%% file: ms.tex
\documentclass[]{aastex631}
\usepackage{color}

\shorttitle{Orbital Architecture of Planetary Systems}
\shortauthors{Kokubo et al.}

\graphicspath{{./}{figures/}}

\newcommand{\scale}[3]{\left(\frac{#1}{#2}\right)^{#3}}
\newcommand{\au}{\mathrm{au}}

\renewcommand{\baselinestretch}{1.3}

\begin{document}

%\title{Orbital Architecture of Planetary Systems Formed by Gravitational Scattering and Collisions}
\title{A Scaling Law for the Orbital Architecture of Planetary Systems \\ Formed by Gravitational Scattering and Collisions}

\author[0000-0002-5486-7828]{Eiichiro Kokubo}
\affiliation{Division of Science, National Astronomical Observatory of Japan, Osawa, Mitaka, Tokyo 181-8588, Japan}
\affiliation{Center for Computational Astrophysics, National Astronomical Observatory of Japan, Osawa, Mitaka, Tokyo 181-8588, Japan}
\affiliation{Department of Astronomy, University of Tokyo, Hongo, Bunkyo, Tokyo 113-0033, Japan}
\email{kokubo.eiichiro@nao.ac.jp}

\author[0000-0001-5953-3897]{Haruka Hoshino}
\affiliation{Department of Astronomy, University of Tokyo, Hongo, Bunkyo, Tokyo 113-0033, Japan}
\affiliation{Division of Science, National Astronomical Observatory of Japan, Osawa, Mitaka, Tokyo 181-8588, Japan}

\author[0000-0002-2383-1216]{Yuji Matsumoto}
\affiliation{Center for Computational Astrophysics, National Astronomical Observatory of Japan, Osawa, Mitaka, Tokyo 181-8588, Japan}

\author[0000-0002-1084-3656]{Re'em Sari}
\affiliation{Racah Institute of Physics, Hebrew University of Jerusalem, 9190401, Israel}

\input{abstract}

\input{introduction}

\input{method}

\input{result}

\input{summary}

\input{acknowledgement}

\bibliographystyle{aasjournal}
\bibliography{myreference}

\end{document}

%% file: abstract.tex
% < 250 words

\begin{abstract}

In the standard formation models of terrestrial planets in the solar system and close-in super-Earths in non-resonant orbits recently discovered by exoplanet observations, planets are formed by giant impacts of protoplanets or planetary embryos after the dispersal of protoplanetary disk gas in the final stage. 
This study aims to theoretically clarify a fundamental scaling law for the orbital architecture of planetary systems formed by giant impacts.
In the giant impact stage, protoplanets gravitationally scatter and collide with one another to form planets. 
Using {\em N}-body simulations, we investigate the orbital architecture of planetary systems formed from protoplanet systems by giant impacts. 
As the orbital architecture parameters, we focus on the mean orbital separation between two adjacent planets and the mean orbital eccentricity of planets in a planetary system. 
%We find that the orbital architecture is scaled by the eccentric distance or the epicycle amplitude for the eccentricity given by the ratio of the planets' two-body surface escape velocity to the Keplerian circular velocity. 
We find that the orbital architecture is determined by the ratio of the two-body surface escape velocity of planets $v_\mathrm{esc}$ to the Keplerian circular velocity $v_\mathrm{K}$, $k = v_\mathrm{esc}/v_\mathrm{K}$.
The mean orbital separation and eccentricity are about $2 ka$ and $0.3 k$, respectively, where $a$ is the system semimajor axis.
With this scaling, the orbital architecture parameters of planetary systems are nearly independent of their total mass and semimajor axis.

\end{abstract}

%\keywords{planets and satellites: formation, terrestrial planets --- methods: {\em N}-body simulations}

%% file: introduction.tex
% PPVII
% e ~ 0.02-0.05
% i ~ 1-2 (deg)
% Ford スライド確認

\section{INTRODUCTION}
\label{sec:introduction}

% terrestrial planets in the solar system
% observations of close-in terrestrial planets

Current exoplanet surveys reveal that many planets in the Galaxy are close-in planets with semimajor axes smaller than about 0.3 au and masses smaller than about 30 Earth masses \cite[e.g.,][]{2011arXiv1109.2497M,2012ApJS..201...15H, 2017AJ....154..107P}.
%E: 0.3 au -> 0.4 au?
Their composition is not yet well constrained, but most may consist mainly of solid components. 
In the present paper, we call these planets close-in super-Earths.
For general characteristics of the planets, refer to \cite{2023ASPC..534..863W}.
Many of these exoplanets reside in multiplanet systems \citep[e.g.,][]{2013ApJS..204...24B}, and some studies estimate that the typical multiplicity of close-in super-Earths is around three \citep[e.g.,][]{2018ApJ...860..101Z}.
% Kepler results (Fabrycky, Xie, Weiss, ...) (Mulders, He 20200609)
The distribution of period ratios between pairs of adjacent planets roughly follows a log-normal distribution with slight excess around some low-order mean motion resonances \citep[e.g.,][]{2013ApJ...778....7B,2015ARA&A..53..409W}.
The orbital separation between adjacent pairs is in the range of 10-40 $r_\mathrm{H}$ with a peak around 20 $r_\mathrm{H}$, where $r_\mathrm{H}$ is the mutual Hill radius, which is smaller than the mean value of the solar system terrestrial planets, 43 $r_\mathrm{H}$ \citep[e.g.,][]{2014Natur.513..336L,2018AJ....155...48W}.
While the orbits of single planets tend to be more eccentric, those of planets in multiplanet systems tend to have smaller eccentricities and inclinations 
% of order $\sim 0.01$ 
of a few percent \citep[e.g.,][]{2016PNAS..11311431X,2017AJ....154....5H,2020AJ....160..276H}.
In other words, the close-in super-Earth systems are dynamically compact and cold.
In addition, most multiplanet systems show an intrasystem uniformity called the ``peas-in-a-pod'' pattern in which planets have nearly equal radii and regularly spaced, nearly circular, and coplanar orbits \citep[e.g.,][]{2018AJ....155...48W,2021ApJ...920L..34M, 2023ASPC..534..863W}.

% standard formation scenario of solar system terrestrial planets
%The standard formation scenario of terrestrial planets in the solar system consists of three stages: (1) planetesimal formation from dust, (2) protoplanet formation from planetesimals, and (3) terrestrial planet formation from protoplanets \citep[e.g.,][]{2012PTEP.2012aA308K, 2014prpl.conf..595R}.
%Stage (3) is known as the giant impact stage, where protoplanets gravitationally scatter and collide with one another to complete planets and then the orbital structure is determined.
It is widely accepted that the final stage of terrestrial planet formation in the solar system involves giant impacts among protoplanets or planetary embryos \citep[e.g.,][]{2012PTEP.2012aA308K, 2014prpl.conf..595R}.
In this stage, protoplanets gravitationally scatter and collide with one another to form planets, and then the orbital structure is determined.
% formation scenarios of close-in systems
%For the formation of close-in terrestrial planets, three models have been proposed: (1) in-situ formation, (2) accretion and then migration and (3) migration and then accretion.
%The formation models of close-in super-Earths can be divided into two categories: ``in-situ completion'' and ``ex-situ completion''.
%The in-situ formation model is the extension of the standard formation scenario of solar system terrestrial planets to inner massive disks \citep[e.g.,][]{}. 
%In the in-situ completion model, planets are formed through giant impacts after disk gas dispersal from a protoplanet system that are formed in situ or by migration from the outer disk \citep[e.g.,][]{2012ApJ...751..158H,2016ApJ...817...90L,2018A&A...615A..63O,2020A&A...642A..23M}. 
The same is assumed for some formation models of the super-Earth systems, where planets are formed through giant impacts after the dispersal of disk gas from a protoplanet system that are formed in situ or by migration from the outer disk \citep[e.g.,][]{2012ApJ...751..158H,2016ApJ...817...90L,2018A&A...615A..63O,2020A&A...642A..23M}. 
Although the formation timescale of protoplanets may be shorter than the lifetime of a gas disk, the presence of gas imposes dynamical friction that circularizes their orbits and prevents their mutual collisions. 
The giant impact phase is then delayed until most of the gas leaves the system.
One of this model's merits is the natural formation of planets in non-resonant orbits.
%In the acretion and then migration model, planets are first formed in the outer disk and then migrate inward due to planet-gas interaction \citep[e.g.,][]{}.
%In contrast, in the ex-situ completion model, planets are first formed in the outer disk and then migrate inward due to planet-gas interaction, which often results in the formation of planets in resonant chains \citep[e.g.,][]{2012MNRAS.427L..21R,2012ARA&A..50..211K}.
%On the other hand, the migration and then accretion model assumes that planet-building materials such as dust, planetesimals, and protoplanets migrate first and then from them terrestrial planets are formed \citep[e.g.,][]{}.
%In models (1) and (2) the final formation stage is the giant impact.
%Therefore the dynamical and accretionary evolution in the giant impact stage is crucial to understand the formation of close-in planets. 
%In this paper, we focus on forming planetary systems by giant impacts that may be responsible for creating the majority of close-in super-Earths.
%While both scenarios may be taking place in different systems, in this paper, we focus on the in-situ completion model that may be responsible for creating the majority of close-in super-Earths. 
In this paper, we focus on the general dynamics of the giant impact evolution that may be responsible for creating the non-resonant close-in super-Earth systems. 
We therefore examine an idealized model of giant impacts, starting with nearly circular, slightly inclined orbits without disk gas.

% previous studies (Chiang, Lee, Dawson, ...)
There are many studies on the giant impact formation of terrestrial planets in the solar system and super-Earths \citep[e.g.,][]{1998Icar..136..304C,2006ApJ...642.1131K, 2018A&A...615A..63O}.
However, only a few papers focused on the orbital architecture of multiplanet systems, such as the mean orbital separation between adjacent planets and the mean orbital eccentricity of planets in a system.
As one of the few examples, \cite{2023MNRAS.519.2838H} investigated the effects of stellar mass on orbital architecture and found that a planetary system becomes dynamically more compact and colder as the stellar mass increases.
% MK+?, Dawson+16
% 軌道構造に注目した系統的なシミュレーション研究はあまりない。
As a related study, \cite{2024MNRAS.527...79G} also examined the effect of dynamical instability on the orbital architecture of multiplanet systems.

% goals 
This study aims to obtain a general scaling law for the orbital architecture of planetary systems formed by giant impacts.
%In the present paper, we investigate the orbital architecture of
% planetary systems formed by giant impacts.
We perform {\em N}-body simulations of the giant impact stage from protoplanet systems.
We clarify how a protoplanet system evolves through gravitational scattering and collisions among protoplanets and what determines the final system configuration.
%and finally derive the basic scaling law of the orbital architecture.

% plan of development

We describe the calculation method in Section~\ref{sec:method} and present the results in Section~\ref{sec:result}.
Section~\ref{sec:summary} is devoted to a summary and discussion.

%% file: method.tex
\section{CALCULATION METHOD}
\label{sec:method}

\subsection{Initial Protoplanet Systems}

\begin{deluxetable}{rrrrrrrr}
\tablewidth{0pt}
\tablecaption{Initial Conditions of Protoplanet Systems \label{tab:initial}}
\tablehead{
 \colhead{model} &
 \colhead{$\Sigma_1\,(\mathrm{gcm}^{-2})$} &
 \colhead{$r_\mathrm{in}\,(\mathrm{au})$} &
 \colhead{$r_\mathrm{out}\,(\mathrm{au})$} &
  \colhead{$\sigma_e$} & 
 \colhead{$N$} & \colhead{$M_\mathrm{tot}\,(M_\oplus)$} &
 \colhead{$a_M\,(\mathrm{au})$}
}
\startdata
S0 & 10 & 0.1 & 0.3 & 0.01 & 15 & 2.44 & 0.17 \\
M1 & 5 & & & 0.005 & 22 & 1.26 & 0.18 \\
M2 & 20 & & & 0.014 & 11 & 5.06 & 0.18 \\
M3 & 50 & & & 0.022 & 7 & 12.75 & 0.18 \\
R1 & & 0.05 & 0.15 & & & & 0.09 \\ 
R2 & & 0.2 & 0.6 & & & & 0.35 \\
R3 & & 0.5 & 1.5 & & & & 0.87 \\
\enddata
\tablecomments{
Blank spaces assume an identical parameter as the standard model S0.
}
\end{deluxetable}
%E: a_M 必要?

% disk model

We perform $N$-body simulations of the giant impact stage from protoplanet systems.
We adopt idealized protoplanet systems to facilitate understanding of the elementary process of orbit determination by giant impacts and its dependence on system parameters.
For simplicity, we assume that the global mass distribution of protoplanets obeys the solid surface density distribution given by
\begin{equation}
 \Sigma = \Sigma_1\left(\frac{r}{1\, \au}\right)^{-2}, 
\end{equation}
 with inner and outer cutoffs, $r_\mathrm{in}$ and $r_\mathrm{out}$, where $\Sigma_1$ is the reference surface density at 1 au, and $r$ is the radial distance from the central star.
Since the giant impact stage starts after the dispersal of disk gas, we assume a gas-free disk.
We systematically vary the disk parameters: $\Sigma_1 = 5, 10, 20, 50\,\mathrm{gcm}^{-2}$, and $(r_\mathrm{in}, r_\mathrm{out}) = (0.05, 0.15), (0.1, 0.3), (0.2, 0.6), (0.5, 1.5)$ (the unit is au).
%E: (r_in, r_out)) = (0.02, 0.06)?
%Note that disks of $\Sigma_1 = 10$ and 50 with $\alpha = 3/2$ correspond to the minimum-mass solar nebula \citep{1977Ap&SS..51..153W, 1981PThPS..70...35H} and extrasolar nebula models \citep{2013MNRAS.431.3444C, 2020AJ....159..247D}, respectively.
We set the central star's mass equal to the solar mass $M_* = M_\odot$.

% protoplanet model
 
%In the oligarchic growth model, the mass of a protoplanet $M$ at the semimajor axis $a$ is given by the isolation mass
The isolation mass gives the protoplanet mass $M$
\begin{equation} 
 \label{eq:m_iso}
 M_\mathrm{iso}
 \simeq 2\pi ab\Sigma = 
 0.16
% \scale{\tilde{b}_\mathrm{H}}{10}{3/2}
 \scale{\Sigma_1}{10}{3/2}
% \scale{a}{1\, \au}{(3/2)(2-\alpha)}
% \scale{M_*}{M_\odot}{-1/2} M_\oplus,
M_\oplus,
\end{equation}
where $a$ is the semimajor axis of a protoplanet, $b$ is the orbital separation between adjacent protoplanets,
%where $b$ is the orbital separation between adjacent protoplanets, $\tilde{b} = b/r_\mathrm{H}$, $r_\mathrm{H}$ is the Hill radius $r_\mathrm{H} = (2M_\mathrm{iso}/3M_\odot)^{1/3}a$ of the protoplanet, $M_*$ is the mass of the central star, and $M_*$ and $M_\oplus$ are solar and Earth masses, respectively.
and $M_\oplus$ is the Earth mass, and we assume $b = 10 r_\mathrm{H}$ \citep[][]{1998Icar..131..171K, 2002ApJ...581..666K}.
The mutual Hill radius for protoplanets $j$ and $j+1$ is given by
\begin{equation}
 r_{\mathrm{H},j} = h_j \frac{a_j + a_{j+1}}{2},
\end{equation}
where $h_j$ is given by
\begin{equation}
 h_j = \left(\frac{M_j + M_{j+1}}{3 M_*} \right)^{1/3},
\end{equation}
where protoplanet numbers $j$ are sorted in ascending order by $a$.
Note that the $\Sigma \propto r^{-2}$ disk corresponds to an equal-mass protoplanet system.
%We vary the orbital parameters of protoplanets to investigate their effects on the final planetary systems.
%The orbital separation of protoplanets is set as $\tilde{b}_\mathrm{H} = 8, 10, 12$, where 10 is the typical value in $N$-body simulations \citep[e.g.,][]{2000Icar..143...15K,2002ApJ...581..666K}. 
%E: b = 6?
The initial eccentricities $e$ and inclinations $i$ of protoplanets are given by the Rayleigh distribution with dispersions
%  $\langle e^2\rangle^{1/2} = 2\langle i^2\rangle^{1/2} = 0.01 (\Sigma_1/10)^{1/2}$
$\sigma_e = 2\sigma_i = 0.01 (\Sigma_1/10)^{1/2}$ (the unit of $i$ is radian).
%E: describe e~ value for alpha = 2 systems
%We also adopt 1/8, 1/4, 1/2, 2, 4, and 8 times the above $\sigma_e$.
%We also adopt 1/4 and 4 times the above $\sigma_e$.
We assume spherical protoplanets with the bulk density $\rho = 3$ gcm$^{-3}$.
%We assume spherical protoplanets with the bulk density $\rho = 3$ gcm$^{-3}$ in most systems with exceptions 1/8 and 8 times different $\rho$ in two models to check the effect of the bulk density.
The initial conditions of protoplanet systems ($\Sigma_1$, $r_\mathrm{in}$, $r_\mathrm{out}$, $\sigma_e$) are summarized in Table~\ref{tab:initial} with their system properties: protoplanet number $N$, total mass $M_\mathrm{tot}$, and mass-weighted mean semimajor axis $a_M = \sum_j^N M_j a_j /\sum_j^N M_j$.
We refer to model S0 as the standard model.
We perform 20 runs for each model with different initial angular distributions of protoplanets.

\subsection{Time Evolution}

% orbital integration

The orbits of protoplanets are calculated by numerically integrating the equation of motion of protoplanets.
%We set the mass of the central star equal to the solar mass.
For numerical integration, we use the modified Hermite integrator for planetary $N$-body simulation \citep{1998MNRAS.297.1067K,2004PASJ...56..861K} with the hierarchical timestep \citep{1991PASJ...43..859M}. 
We use the Phantom GRAPE scheme \citep{2006NewA...12..169N} to efficiently compute mutual gravity among protoplanets.
The simulations follow the evolution of protoplanet systems for at least $5\times 10^8 t_\mathrm{K}$ of the innermost protoplanet, where $t_\mathrm{K}$ is the Kepler period.

% accretion

During orbital integration, when two protoplanets contact, a collision occurs.
For simplicity, we adopt perfect accretion as the collision model, where every collision leads to accretion.
The dynamical properties of planets formed by giant impacts barely change even if hit-and-run collisions are considered \citep{2010ApJ...714L..21K}.

\subsection{System Parameters}
\label{sec:system_parameter}

%E: add solar system value?

We investigate the properties of not individual planets but a planetary system as a whole.
We introduce the architecture parameters normalized by the mutual Hill radius, which are the mean orbital separation between adjacent planets $j$ and $j+1$
\begin{equation}
 \tilde{b}_\mathrm{H} = \frac{1}{N -1} \sum_j^{N-1} \frac{a_{j+1} - a_j}{r_{\mathrm{H},j}},
\label{eq:bH}
\end{equation}
and the mean eccentricity (epicycle amplitude)
\begin{equation}
 \tilde{e}_\mathrm{H} = \frac{1}{N-1} \sum_j^{N-1} \frac{a_j e_j + a_{j+1}e_{j+1}}{2 r_{\mathrm{H},j}}.
\label{eq:eH}
\end{equation}
We calculate $\tilde{i}_\mathrm{H}$ in the same way.
%The mutual Hill radius for planets $j$ and $j+1$ is given by
%\begin{equation}
% r_{\mathrm{H},j} = h_j \frac{a_j + a_{j+1}}{2},
%\end{equation}
%where $h_j$ is given by
%\begin{equation}
% h_j = \left(\frac{M_j + M_{j+1}}{3 M_*} \right)^{1/3}.
%\end{equation}
%where $M_*$ is the stellar mass and $M$ and $a$ are the mass and semimajor axis of planets, respectively.
%E: solar system terrestrial planets
Note that with this definition $\tilde{e}_\mathrm{H}$ is addable to $\tilde{b}_\mathrm{H}$. 
To measure the separation uniformity in a system, we calculate the standard deviation $\sigma_{\tilde{b}_\mathrm{H}}$ of $\tilde{b}_\mathrm{H}$, and inspect the normalized separation deviation 
$\sigma_{\tilde{b}_\mathrm{H}}/\tilde{b}_\mathrm{H}$.
%As the system semimajor axis, we calculate the mass-weighted mean semimajor axis of planets, $a_M = \sum_j^N M_j a_j /\sum_j^N M_j$.
As the system semimajor axis, we use the mass-weighted mean semimajor axis of planets $a_M$.
For example, the terrestrial planet system in the solar system has $\tilde{b}_\mathrm{H} = 43$, $\tilde{e}_\mathrm{H} = 5$, $\sigma_{\tilde{b}_\mathrm{H}}/\tilde{b}_\mathrm{H} = 0.35$ and $a_M = 0.9\,\au$.

In addition, as the physical properties of a planetary system, we calculate the number $N$, maximum mass $M_\mathrm{max}$, average mass $M_\mathrm{ave}$, and mass ratio between standard deviation and average $\sigma_M/M_\mathrm{ave}$ of planets.
Note that the number of planets per unit logarithmic semimajor axis is given by $N/\log(r_\mathrm{out}/r_\mathrm{in}) = N/\log3$.
%For example, the solar system terrestrial planets have ...

We calculate the median and median absolute deviation of these parameters in 20 runs for each model, which are given with $\langle\cdot\rangle$.

%% file: result.tex
% Figures
%   fit lines?
%   unify label spaces
% Ghosh, T., & Chatterjee, S. 2024, MNRAS, 527, 79, doi: 10.1093/mnras/stad2962

\section{RESULTS}
\label{sec:result}

\begin{deluxetable}{rrrrrrrrrrrr}
\tabletypesize{\scriptsize}
\tablewidth{0pt}
\tablecaption{Physical and Orbital Properties of Planetary Systems \label{tab:mass_orbit}}
\tablehead{
 \colhead{model} &
 \colhead{$\langle N\rangle$} &
 \colhead{$\langle M_\mathrm{max}\rangle$\,($M_\oplus$)} & 
 \colhead{$\langle M_\mathrm{ave}\rangle$\,($M_\oplus$)} & 
 \colhead{$\langle \sigma_M/M_\mathrm{ave}\rangle$} &
 \colhead{$\langle\tilde{b}_\mathrm{H}\rangle$} &
 \colhead{$\langle\tilde{e}_\mathrm{H}\rangle$} & 
 \colhead{$\langle\tilde{i}_\mathrm{H}\rangle$} & 
 \colhead{$\langle\tilde{b}_\mathrm{K}\rangle$} &
 \colhead{$\langle\tilde{e}_\mathrm{K}\rangle$} &
 \colhead{$\langle\tilde{i}_\mathrm{K}\rangle$} &
 \colhead{$\langle \sigma_{\tilde{b}_\mathrm{H}}/\tilde{b}_\mathrm{H}\rangle$} 
}
\startdata
S0 & $5.0 \pm 1.0$ & $0.81 \pm 0.16$ & $0.49 \pm 0.08$ & $0.38 \pm 0.09$ & $24.56 \pm  2.46$ & $3.10 \pm 0.70$ & $1.87 \pm 0.57$ & $2.16 \pm 0.20$ & $0.29 \pm 0.05$ & $0.16 \pm 0.04$ & $0.14 \pm 0.06$ \\
M1 & $6.0 \pm 1.0$ & $0.29 \pm 0.00$ & $0.21 \pm 0.03$ & $0.27 \pm 0.04$ & $24.78 \pm  2.69$ & $3.17 \pm 0.60$ & $2.16 \pm 0.69$ & $2.26 \pm 0.24$ & $0.28 \pm 0.06$ & $0.18 \pm 0.07$ & $0.13 \pm 0.03$ \\
M2 & $4.0\pm 1.0$ & $1.83 \pm 0.46$ & $1.26 \pm 0.25$ & $0.26 \pm 0.10$ & $23.75 \pm 3.34$ & $2.77 \pm 0.74$ & $2.07 \pm 0.90$ & $2.12 \pm 0.25$ & $0.25 \pm 0.07$ & $0.18 \pm 0.08$ & $0.15 \pm 0.07$ \\
M3 & $3.0\pm 1.0$ & $5.44 \pm 1.80$ & $4.23 \pm 1.10$ & $0.38 \pm 0.13$ & $20.18 \pm  3.94$ & $2.64 \pm 0.77$ & $1.37 \pm 0.48$ & $1.81 \pm 0.29$ & $0.23 \pm 0.08$ & $0.12 \pm 0.04$ & $0.08 \pm 0.05$ \\
R1 & $6.0 \pm 0.5$ & $0.65 \pm 0.00$ & $0.41 \pm 0.03$ & $0.34 \pm 0.05$ & $20.92 \pm 1.52$ & $2.36 \pm 0.45$ & $1.44 \pm 0.58$ & $2.66 \pm 0.18$ & $0.31 \pm 0.07$ & $0.17 \pm 0.07$ &
$0.16 \pm 0.02$ \\
R2 & $3.5\pm 0.5$ & $0.97 \pm 0.16$ & $0.71 \pm 0.10$ & $0.39 \pm 0.15$ & $34.31 \pm  2.89$ & $5.54 \pm 1.96$ & $3.84 \pm 0.95$ & $2.10 \pm 0.20$ & $0.35 \pm 0.12$ & $0.23 \pm 0.04$ & $0.14 \pm 0.04$ \\
R3 & $3.0\pm 0.5$ & $1.46 \pm 0.16$ & $0.70 \pm 0.09$ & $0.77 \pm 0.15$ & $54.40 \pm  8.91$ & $11.98 \pm 3.00$ & $10.36 \pm 3.58$ & $1.97 \pm 0.29$ & $0.43 \pm 0.10$ & $0.36 \pm 0.11$ & $0.11 \pm 0.10$ 
\enddata
%\tablecomments{}
\end{deluxetable}

We investigate the orbital architecture of planetary systems formed from protoplanet systems by giant impacts.
Firstly, we show the time evolution of the orbital architecture from protoplanet systems to planetary systems in Section \ref{sec:time_evolution}.
We next examine how the orbital architecture depends on the system mass and semimajor axis in Sections \ref{sec:m-dependence} and \ref{sec:a-dependence}.
Then, we derive a scaling law of the orbital architecture in Section \ref{sec:new_scaling}.
The statistical results of the physical and orbital properties of all models are summarized in Table~\ref{tab:mass_orbit}. 
%In the following, we explore some of the main results.
%The physical properties are the number $N$, maximum mass $M_\mathrm{max}$, average mass $M_\mathrm{ave}$, and mass ratio between standard deviation and average $\sigma_M/M_\mathrm{ave}$.
%E: M_max/ave/M_tot にする?

\subsection{Time Evolution}
\label{sec:time_evolution}

%Firstly, to illustrate the simulation and analysis, we show the results of the standard model (model S0). 

% Figure: t-<N>,<~b>,<~e>,<~i> (t-nbei.mon)
\begin{figure*}
\centering
\includegraphics[width=0.5\textwidth]{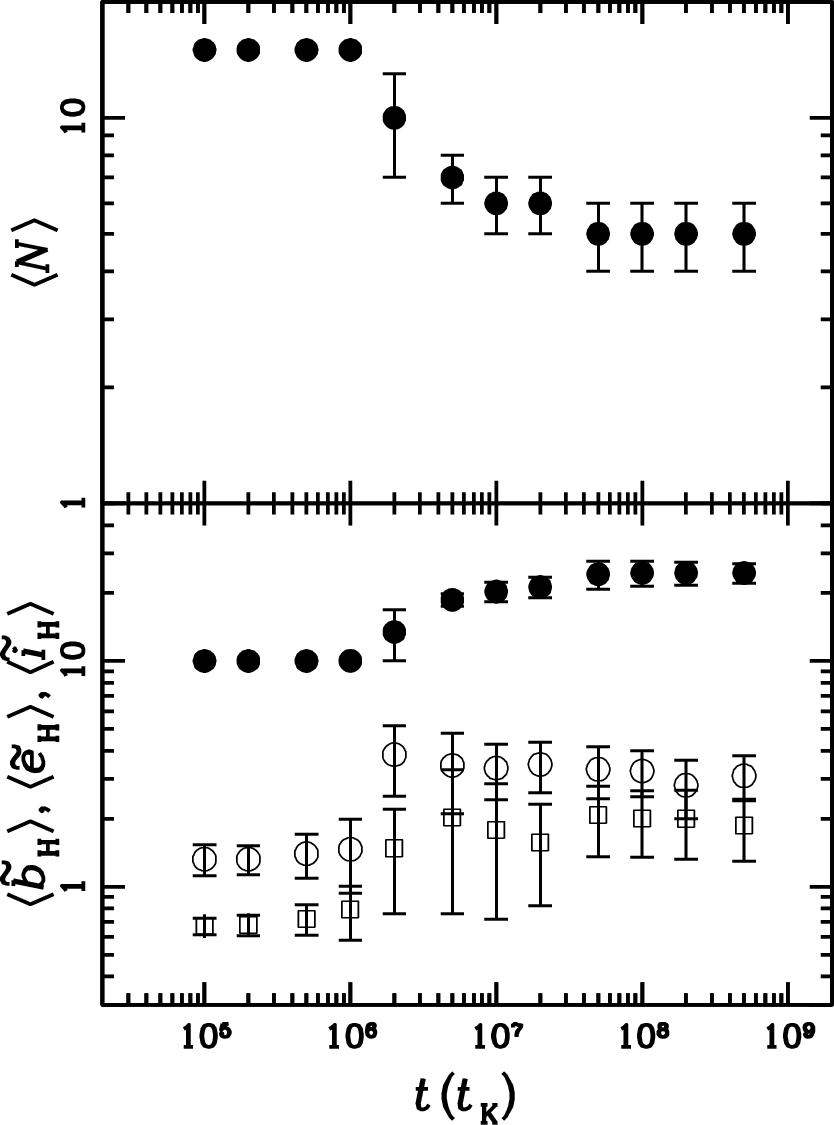}
\caption{
Time evolution of the number $\langle N\rangle$ (top panel) and the normalized orbital separation between adjacent pairs $\langle\tilde{b}_\mathrm{H}\rangle$ (filled circles), eccentricity $\langle\tilde{e}_\mathrm{H}\rangle$ (open circles), and inclination $\langle\tilde{i}_\mathrm{H}\rangle$ (open squares) (bottom panel) of the standard protoplanet systems (model S0).
The values and error bars indicate the median and median absolute deviation in 20 runs, respectively.
}
\label{fig:t-nbei}
\end{figure*}

Firstly, we show the time evolution of the orbital architecture using the standard model (model S0). 
The averaged number of planets and scaled orbital architecture calculated from 20 runs are plotted against time in Figure~\ref{fig:t-nbei}.
The system is quite stable for about $10^6\, t_\mathrm{K}$ in the sense that the orbital architecture shows no substantial changes. 
None of the systems exhibit a merger before $10^6\, t_\mathrm{K}$, while several mergers typically occur before $2\times 10^6\, t_\mathrm{K}$, reducing the number of bodies on average by $1/3$. 
%Initially, the system is stable for about $10^6\, t_\mathrm{K}$ in the sense that the orbital architecture shows no substantial changes. 
%Then, it becomes unstable, and the giant impacts begin.
The giant impact stage lasts for several $10^7\, t_\mathrm{K}$ forming planetary systems, which are stable for longer than $10^8\, t_\mathrm{K}$. 
This evolution is seen in the number of planets $\langle N\rangle$ in Figure~\ref{fig:t-nbei}.
%Over the whole evolution, the mass-weighted semimajor axis $\langle a_M \rangle$ is kept constant. 
%In the giant impact stage generally the mass-weighted eccentricity $\langle e_M \rangle$ and inclination $\langle i_M \rangle$ and thus the normalized AMD $\langle D \rangle$ are pumped up by gravitational scattering, while they are damped by collisions.
%The orbital (in)stability determines the final orbital architecture. 
%Figure~\ref{fig:t-oa} shows the scaled orbital architecture parameters $\langle\tilde{e}_\mathrm{H}\rangle$, $\langle\tilde{i}_\mathrm{H}\rangle$, and $\langle\tilde{b}_\mathrm{H}\rangle$.
The initial parameters of the orbital architecture are $\tilde{b}_\mathrm{H} = 10$ and $\tilde{e}_\mathrm{H} \simeq 2\tilde{i}_\mathrm{H} \simeq 1.3$ in this model.
We find that when and after the giant impact stage begins, $\langle\tilde{e}_\mathrm{H}\rangle$ and $\langle\tilde{i}_\mathrm{H}\rangle$ increase rapidly and then become almost constant with time, while $\langle\tilde{b}_\mathrm{H}\rangle$ increases monotonically with time. 
This is the typical evolution of the scaled orbital architecture by giant impacts. 
The final parameters of the orbital architecture are $\langle\tilde{b}_\mathrm{H}\rangle \simeq 25$, $\langle\tilde{e}_\mathrm{H}\rangle \simeq 3.1$, and $\langle\tilde{i}_\mathrm{H}\rangle \simeq 1.9$, which are determined by the orbital (in)stability. 
%The final orbital architecture is determined by the orbital (in)stability. 
%Since the behavior of the inclination is similar to that of the eccentricity in the present parameter range, we focus on the eccentricity hereafter.

%Using the results of 20 runs we obtain the statistical results of this model, which are given in Tables~\ref{tab:mass} and \ref{tab:orbit}.
%We find that compared with the previous works for the terrestrial planet system in the solar system \citep[e.g.,][]{2006ApJ...642.1131K}, a more compact, dynamically colder planetary system is formed.
We find that compared to previous works that investigated giant impact evolution with the larger semimajor axis ($a \sim 1\,\mathrm{au}$) \citep[e.g.,][]{2006ApJ...642.1131K}, a more compact, dynamically colder planetary system is formed.
The origin of this difference by the semimajor axis will be discussed in detail in Section~\ref{sec:a-dependence}.
As could be expected from giant impacts, the planet's orbits in the final systems are basically not in mean motion resonances, consistent with most observed Kepler planets \citep[e.g.,][]{2015ARA&A..53..409W}.
The orbital separation is also consistent with the mode of the orbital separation distribution of Kepler multiplanet systems in the same region \citep[e.g.,][]{2014Natur.513..336L}.
The ratio between the mean eccentricity and inclination is $\langle \tilde{i}_\mathrm{H}\rangle/\langle \tilde{e}_\mathrm{H}\rangle \sim 0.5$, which is typical for planetary systems formed by giant impacts \citep[e.g.,][]{2006ApJ...642.1131K,2017AJ....154...27M}.
Since the inclination's behavior is similar to the eccentricity in the present parameter range, we focus on the eccentricity hereafter.

We find that the orbital architecture parameters converge to certain values.
The next step is to examine how these values depend on the system parameters.

\subsection{Dependence on System Mass}
\label{sec:m-dependence}

% Figure: M_tot-m,b,e (m-mbe2.mon)
\begin{figure*}
\centering
\includegraphics[width=0.5\textwidth]{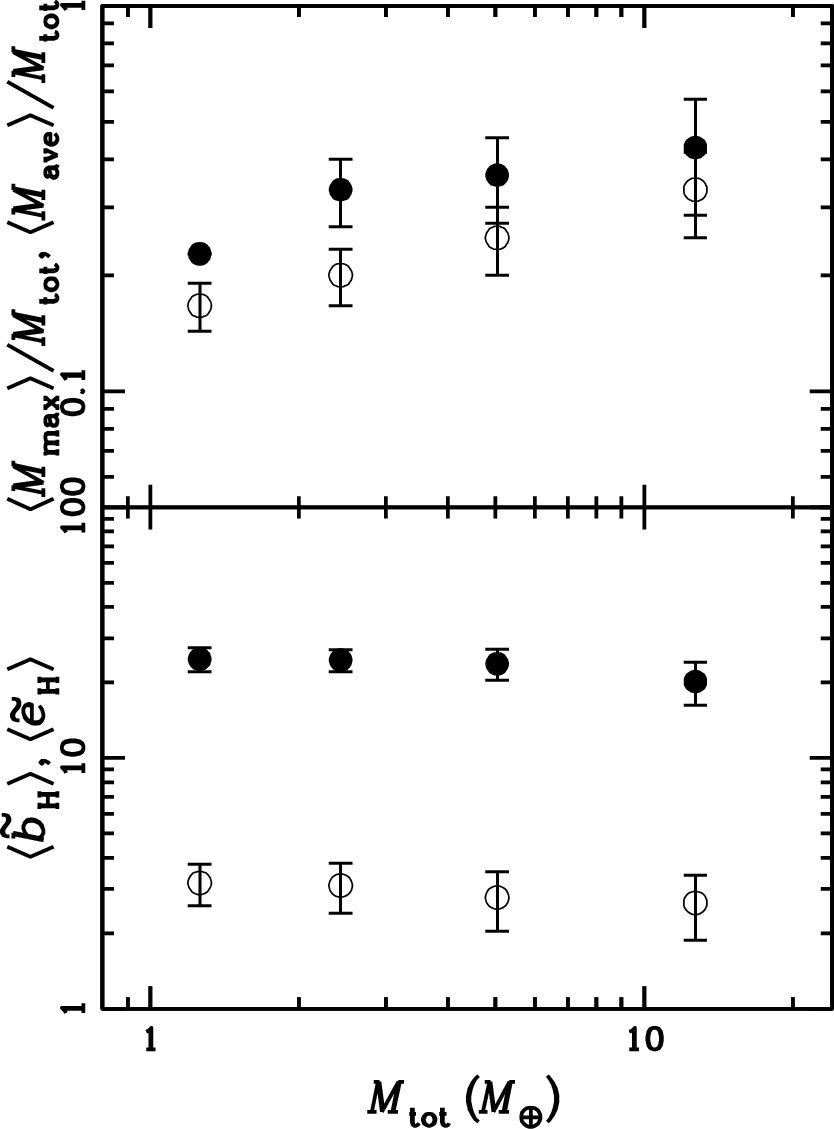}
\caption{
Maximum $\langle M_\mathrm{max}\rangle$ (filled circles) and average $\langle M_\mathrm{ave}\rangle$ (open circles) masses of planets normalized by the system total mass $M_\mathrm{tot}$ (top panel) and the normalized orbital separation between adjacent pairs $\langle\tilde{b}_\mathrm{H}\rangle$ (filled circles) and eccentricity $\langle\tilde{e}_\mathrm{H}\rangle$ (open circles) (bottom panel) against the total mass of protoplanet systems with the reference surface density $\Sigma_1 = 5, 10, 20$, and $50$ (models S0 and M1-3).
}
\label{fig:m-mbe}
\end{figure*}
% M_tot       M_max/M_tot  M_ave/M_tot  ~b_H        ~e_H 
% 1.258451343 0.2272730023 0.1666667461 24.78365135 3.173635006
% 2.427610397 0.3333331347 0.2000003904 24.55920029 3.097424984
% 5.038293362 0.3636350334 0.2499993145 23.74955177 2.773725033
% 12.69608784 0.4285706282 0.3333317935 20.17605019 2.641359806%

We investigate the effects of the disk parameter $\Sigma_1$ on the orbital architecture of planetary systems by comparing models with $\Sigma_1 = 5, 10, 20$, and $50$ (models S0 and M1-3). 
Changing $\Sigma_1$ corresponds to changing the individual initial and total protoplanet masses.
The total mass of protoplanets increases in proportion to $\Sigma_1$.
Figure~\ref{fig:m-mbe} shows the maximum and average masses of planets and the normalized orbital separation and eccentricity of planetary systems against the total protoplanet mass $M_\mathrm{tot}$.
As evident from the top panel of Figure~\ref{fig:m-mbe}, $\langle M_\mathrm{max}\rangle$ and $\langle M_\mathrm{ave}\rangle$ increase with $M_\mathrm{tot}$ \citep[][]{2006ApJ...642.1131K}.
Correspondingly, $\langle N\rangle$ decreases with increasing $M_\mathrm{tot}$ as in Table \ref{tab:mass_orbit}.
%The average mass $\langle M_\mathrm{ave}\rangle$ also shows a similar trend.
%These results indicate that the masses are similarly determined for the total mass.
By the least-square-fit method, we obtain $\mathrm{d}\log (\langle M_\mathrm{max}\rangle/M_\mathrm{tot})/\mathrm{d}\log M_\mathrm{tot} \simeq \mathrm{d}\log(\langle M_\mathrm{ave}\rangle/M_\mathrm{tot})/\mathrm{d}\log M_\mathrm{tot}\simeq 0.3$.
The normalized mass deviation $\langle\sigma_M/M_\mathrm{ave}\rangle$ is about 0.3-0.4, which is consistent with \cite{2022AJ....163..201G}.

We also find that both $\tilde{b}_\mathrm{H}$ and $\tilde{e}_\mathrm{H}$ barely depend on $M_\mathrm{tot}$ with $\mathrm{d}\log\langle\tilde{b}_\mathrm{H}\rangle/\mathrm{d}\log M_\mathrm{tot} \simeq \mathrm{d}\log\langle\tilde{e}_\mathrm{H}\rangle/\mathrm{d}\log M_\mathrm{tot}\simeq -0.09$.
In other words, the orbital parameters of planetary systems with different masses can be scaled by the Hill radius.
% 具体的な数字いれる?
The orbital separations in a system for all models are relatively uniform with $\sigma_{\tilde{b}_\mathrm{H}}/\tilde{b}_\mathrm{H} = 0.08$-$0.15$.

\subsection{Dependence on System Semimajor Axis}
\label{sec:a-dependence}

% Figure: a-e-r_in (a-e-r-r_in.mon), a_M-m,b,e (a-mbe4.mon)
\begin{figure*}
\centering
\plottwo{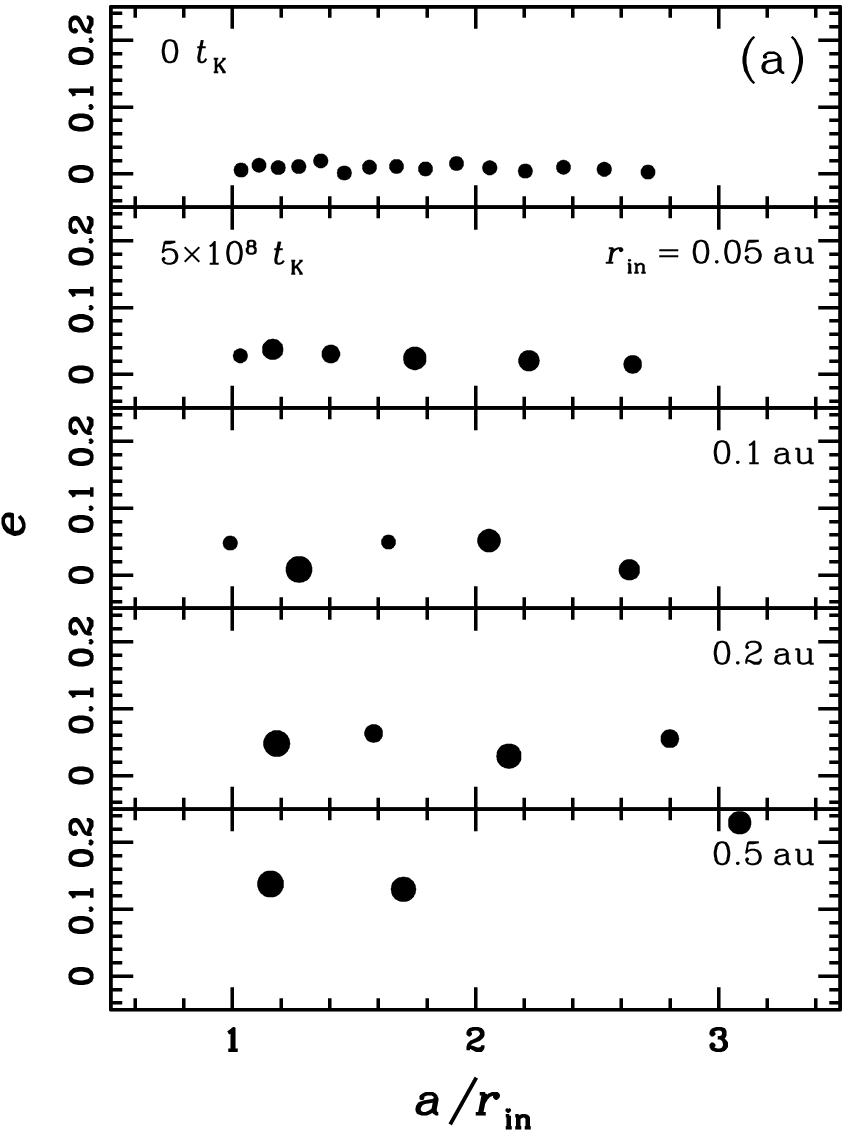}{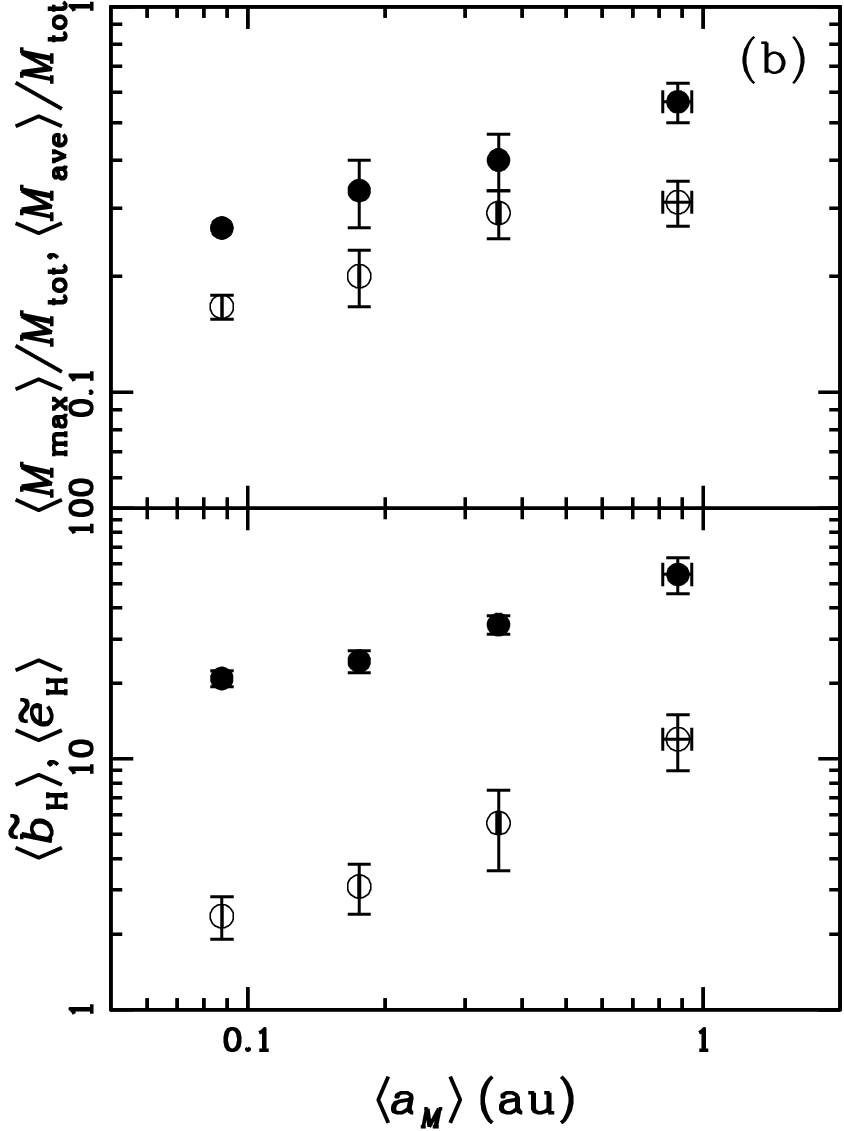}
\caption{
Panel a (left): Examples of the final planetary system on the semimajor-axis $a$-eccentiricity $e$ plane, where $a$ is scaled by the inner edge radius $r_\mathrm{in}$ for protoplanet systems at different locations (models S0 and R1-3), together with the initial protoplanet system (top panel).
The sizes of the circles are proportional to the physical sizes of the planets.
Panel b (right): Maximum $\langle M_\mathrm{max}\rangle$ (filled circles) and average $\langle M_\mathrm{ave}\rangle$ (open circles) masses of planets (top panel) and the orbital separation between adjacent pairs $\langle\tilde{b}_\mathrm{H}\rangle$ (filled circles) and eccentricity $\langle\tilde{e}_\mathrm{H}\rangle$ (open circles) normalized by the Hill radius $r_\mathrm{H}$ (bottom panel) against the system semimajor axis $\langle a_M\rangle$ for $(r_\mathrm{in}, r_\mathrm{out}) = (0.05, 0.15), (0.1, 0.3), (0.2, 0.6)$, and $(0.5, 1.5)$ (the unit is au) (models S0 and R1-3).
}
\label{fig:a_m-dependence}
\end{figure*}
% a_M           M_max        M_ave        ~b_H        ~e_H        ~b_K        ~e_K    
% 0.08761420101 0.2666667998 0.1666669995 20.92134857 2.359534979 2.658864737 0.3134205043
% 0.175530985 0.3333331347 0.2000003904 24.55920029  3.097424984 2.161719799 0.291010499
% 0.3550950289 0.3999995291 0.2916666567 34.31339645 5.544550419 2.10352993 0.3451915085
% 0.8797110319 0.5666683316 0.3111110628 54.40240097 11.98469925 1.970190167 0.4286540151

% Figure: a-e-r_in (a-e-r-r_in.mon)
%\begin{figure*}
%\centering
%\includegraphics[width=0.5\textwidth]{../../new_data/s10/a2/b10/a-e-r-r_in.eps}
%\includegraphics[width=0.5\textwidth]{a-e-r-r_in.eps}
%\caption{
%...
%}
%\label{fig:a-e-r-r_in}
%\end{figure*}

% Figure: a_M-m,b,e (a-mbe2.mon)
%\begin{figure*}
%\centering
%\includegraphics[width=0.5\textwidth]{a-mbe2.eps}
%\caption{
%...
%}
%\label{fig:a-mbe}
%\end{figure*}

Next, we investigate how the orbital architecture of planetary systems depends on the initial location of protoplanet systems (models S0 and R1-3).
Note that with $\Sigma \propto r^{-2}$, $N$ and $M_\mathrm{tot}$ are the same in all these models.

In Figure~\ref{fig:a_m-dependence}a, the final planetary systems of an example run for each model are shown in the $a/r_\mathrm{in}$-$e$ plane, where the semimajor axis is scaled by $r_\mathrm{in}$, and the initial protoplanet systems are the same in this plane.
During formation, the mean semimajor axis of the system is kept almost constant.
For systems located farther away from the star (larger $r_\mathrm{in}$), the number of planets decreases, while the eccentricities and relative orbital separations to the semimajor axis increase.

Figure~\ref{fig:a_m-dependence}b plots $\langle M_\mathrm{max}\rangle$, $\langle M_\mathrm{ave}\rangle$, $\langle\tilde{b}_\mathrm{H}\rangle$ and $\langle\tilde{e}_\mathrm{H}\rangle$ against the system semimajor axis $\langle a_M\rangle$.
We find that both $\langle M_\mathrm{max}\rangle$ and $\langle M_\mathrm{ave}\rangle$ increse with $\langle a_M\rangle$ as well as $\langle\sigma_M/M_\mathrm{ave}\rangle$ in Table \ref{tab:mass_orbit}.
This $\langle\sigma_M/M_\mathrm{ave}\rangle$ dependence indicates that the mass distribution in the inner disk can be more uniform than in the outer disk.
The dependencies of $\langle M_\mathrm{max}\rangle$ and $\langle M_\mathrm{ave}\rangle$ on $\langle a_M\rangle$ are $\mathrm{d}\log (\langle M_\mathrm{max}\rangle/M_\mathrm{tot})/\mathrm{d}\log\langle a_M\rangle \simeq \mathrm{d}\log(\langle M_\mathrm{ave}\rangle/M_\mathrm{tot})/\mathrm{d}\log\langle a_M\rangle \simeq 0.3$.
We find that as $\langle a_M\rangle$ decreases, $\langle\tilde{b}_\mathrm{H}\rangle$ and $\langle\tilde{e}_\mathrm{H}\rangle$ decrease, in other words, the system becomes more compact and dynamically colder in the inner disk.
We obtain $\mathrm{d}\log\langle\tilde{b}_\mathrm{H}\rangle/\mathrm{d}\log\langle a_M\rangle \simeq 0.4$ and $\mathrm{d}\log\langle\tilde{e}_\mathrm{H}\rangle/\mathrm{d}\log\langle a_M\rangle \simeq 0.7$.
This trend is because in the inner disk, collisions are relatively more effective than gravitational scattering since the Hill radius decreases with the semimajor axis, and thus accretion proceeds in a tournament-chart-like way on the $a$-$M$ plane \citep[e.g.,][]{2017AJ....154...27M,2023MNRAS.519.2838H}.
%The resultant orbital separation for $\langle a_M\rangle \simeq 0.2 \au$ is consistent with the observation \citep{}.
The smaller eccentricities realized by collisional damping allow for narrower orbit spacings.
Again, the orbital separation deviation is less than or equal to 0.16 as shown in Table \ref{tab:mass_orbit}, which means rather uniformly spaced orbits.
%Note that the resultant orbital separation for $\langle a_M\rangle \simeq 1\, \au$ (model R3) is almost consistent with that of the solar system terrestrial planets.
Note that the resultant orbital separation $54.40 \pm  8.91$ for $\langle a_M\rangle = 0.87\, \au$ (model R3) is generally consistent with that of the solar system terrestrial planets, 43.

\subsection{New Scaling}
\label{sec:new_scaling}

\begin{figure*}
\centering
\includegraphics[width=0.5\textwidth]{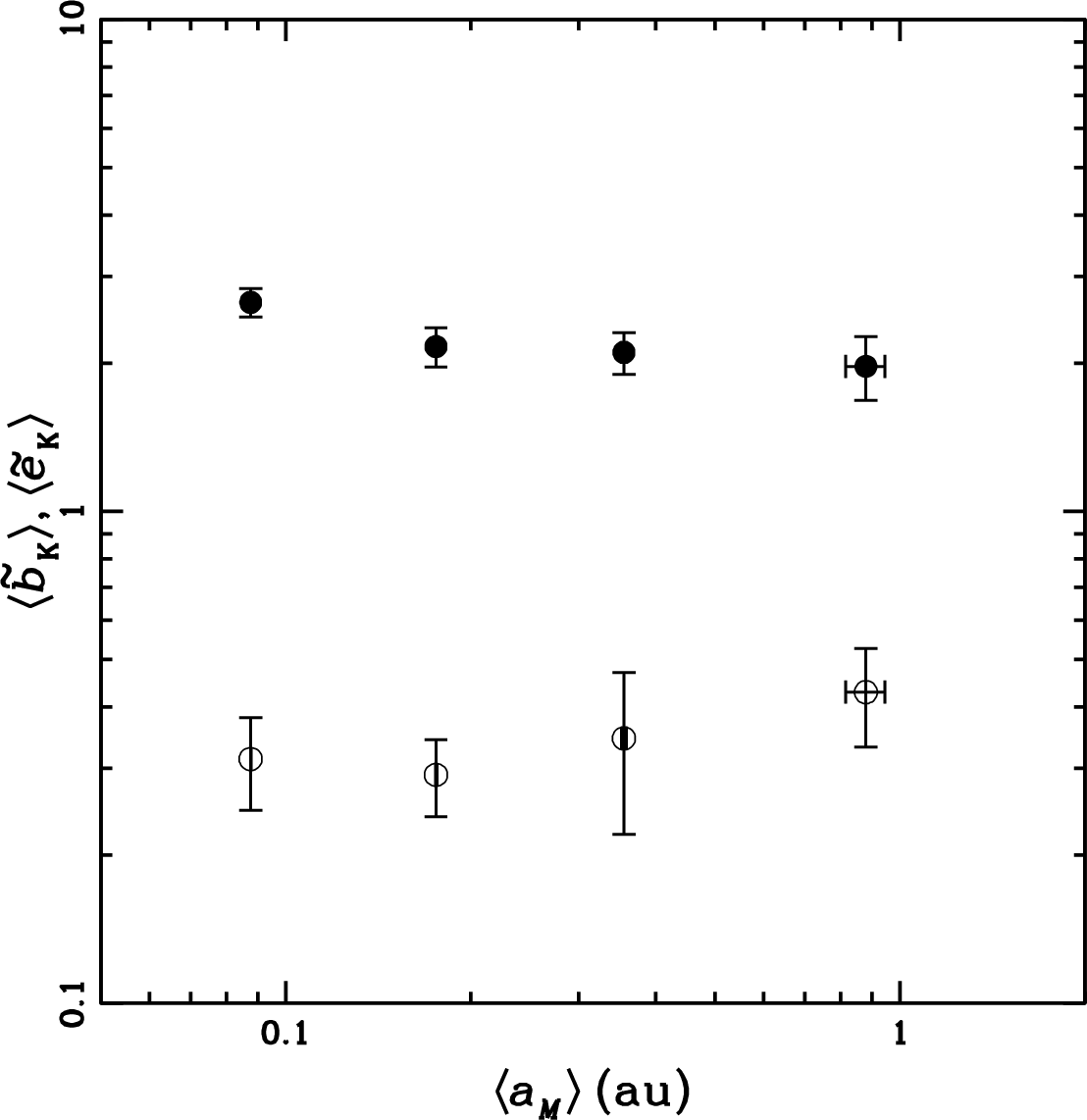}
\caption{
Orbital separation between adjacent pairs $\langle\tilde{b}_\mathrm{K}\rangle$ (filled circles) and eccentricity $\langle\tilde{e}_\mathrm{K}\rangle$ (open circles) normalized by the eccentric distance $r_\mathrm{K}$  against the system semimajor axis $\langle a_M\rangle$ for $(r_\mathrm{in}, r_\mathrm{out}) = (0.05, 0.15), (0.1, 0.3), (0.2, 0.6)$, and $(0.5, 1.5)$ (the unit is au) (models S0 and R1-3).
}
\label{fig:a_m-be_K}
\end{figure*}

The orbital architecture parameters of planetary systems do not depend on the system mass but on the system semimajor axis.
Based on these results, let us consider a scaling law for the orbital architecture.

Here, we introduce a new characteristic length scale that renormalizes the relative importance of gravitational scattering and collisions described above.
When a planetesimal system evolves under gravitational scattering and collisions, the random velocity of planetesimals becomes as large as their two-body surface escape velocity $v_\mathrm{esc}$ \cite[e.g.,][]{1972epcf.book.....S, 1996Icar..123..180K}.
We adopt this idea, although there are fewer bodies in protoplanet systems than in planetesimal systems.
Under this equilibrium, the characteristic eccentricity, often called "escape eccentricity'' is given by $k = v_\mathrm{esc}/v_\mathrm{K}$, where $v_\mathrm{K}$ is the Kepler circular velocity.
Using $k$, we introduce the characteristic length scale $r_\mathrm{K}$ corresponding to the eccentric distance or the amplitude of the radial excursion (epicycle amplitude).
In the same way as the Hill radius, $r_\mathrm{K}$ for planets $j$ and $j+1$ is defined as
\begin{equation}
 r_{\mathrm{K},j} = k_j \frac{a_j + a_{j+1}}{2},
\end{equation}
where
\begin{equation}
 k_j = \frac{v_{\mathrm{esc},j}}{v_{\mathrm{K},j}} = 
 \left[\frac{2G(M_j+M_{j+1})}{R_j+R_{j+1}}\right]^{1/2}
 \left[\frac{GM_*}{(a_j+a_{j+1})/2}\right]^{-1/2} =
%\left(\frac{M_1+M_2}{M_*}\frac{a_1+a_2}{R_1+R_2}\right)^{1/2} =
 h_j \left(6\frac{r_{\mathrm{H},j}}{R_j+R_{j+1}}\right)^{1/2}.
\label{eq:k}
\end{equation}
%E: S(s) -> K(k), G(g), E(e_esc) (重力半径)? 
Note that $r_{\mathrm{K},j}$ consists of the Hill radius $r_{\mathrm{H},j}$ and the Hill-to-physical radius ratio $r_{\mathrm{H},j}/(R_j+R_{j+1})$.
This Hill-to-physical radius ratio $\propto a^{1/2}$ indicates the dependence of the relative importance of gravitational scattering against collisions on the semimajor axis.
We calculate the orbital architecture parameters $\tilde{b}_\mathrm{K}$ and $\tilde{e}_\mathrm{K}$  normalized by $r_{\mathrm{K},j}$ in the same way as
Eqs.(\ref{eq:bH}) and (\ref{eq:eH}), which are given in
Table \ref{tab:mass_orbit}.
%In Figure~\ref{fig:a_m-be_K}, we plot the orbital architecture normalized by $r_{\mathrm{K},j}$.
In Figure~\ref{fig:a_m-be_K}, we plot $\tilde{b}_\mathrm{K}$ and $\tilde{e}_\mathrm{K}$ against $\langle a_M\rangle$ (models S0 and R1-3).
We find that $\langle\tilde{b}_\mathrm{K}\rangle$ and $\langle\tilde{e}_\mathrm{K}\rangle$ are almost independent of $\langle a_M\rangle$ with weak dependencies of $\mathrm{d}\log\langle\tilde{b}_\mathrm{K}\rangle/\mathrm{d}\log\langle a_M\rangle \simeq -0.1$ and $\mathrm{d}\log\langle\tilde{e}_\mathrm{K}\rangle/\mathrm{d}\log\langle a_M\rangle \simeq 0.1$.
This means that the $a_M$-dependence is renormalized in scaling with $r_\mathrm{K}$, which shows that for the orbital architecture, $r_\mathrm{K}$ is a more general scaling unit than $r_\mathrm{H}$.
Note that this $r_\mathrm{K}$ scaling is reduced to the Hill scaling for the fixed Hill-to-physical radius ratio, as in section~\ref{sec:m-dependence}, which again shows $\tilde{b}_\mathrm{K}$ and $\tilde{e}_\mathrm{K}$ independent of $M_\mathrm{tot}$ (models S0 and M1-3) as in Table \ref{tab:mass_orbit}.
%This result is consistent with the characteristic eccentricity in section~\ref{sec:dynamics}. 
%As shown in Eq.~(\ref{eq:k}), $r_\mathrm{K}/r_\mathrm{H} = [(6r_{\mathrm{H},j}/(R_j+R_{j+1})]^{1/2} \propto a^{1/2}$, which is an additional factor responsible for the $a_M$-independence. 
In this new scaling, the orbital architecture is approximately given as $\langle\tilde{b}_\mathrm{K}\rangle \simeq 2.2\pm0.3$ and $\langle\tilde{e}_\mathrm{K}\rangle \simeq 0.3\pm0.1$.
%In this new scaling, the orbital architecture is approximately given by the simple average of the parameters as $\langle\tilde{b}_\mathrm{K}\rangle \simeq 2.2$ and $\langle\tilde{e}_\mathrm{K}\rangle \simeq 0.34$.
%E: 2.2 -> 2, 0.34- > 0.3? 物理的な意味を述べる?
%For an equal-mass system around the solar-mass star, $r_\mathrm{K}$ can be approximated by
For an equal-mass system, $r_\mathrm{K}$ can be approximated by
\begin{equation}
r_\mathrm{K} \simeq ka \simeq 0.011 \scale{M}{M_\oplus}{1/3}\scale{\rho}{3\, \mathrm{gcm}^{-3}}{1/6}\scale{a}{0.1\, \au}{3/2} \scale{M_*}{M_\odot}{-1/2}\, \au.
% acal (32*PI/3)^(1/6)*ME^(1/3)*MS^(-1/2)*3^(1/6)*(0.1*AU)^(3/2)/AU
\end{equation}
%E: 密度いる?中心星質量は?
%E: k の式も出す?

In summary, the above results clearly show that the basic orbital architecture of planetary systems formed by giant impacts is scaled by $r_\mathrm{K}$, in other words, $v_\mathrm{esc}$ determines the spatial structure of planetary systems.
The results of our {\em N}-body simulations, $\tilde{b}_\mathrm{K} \simeq 2$, justify the  assumptions in some planet accretion models \cite[e.g.,][]{2014ApJ...795L..15S,2023ASPC..534..863W}.
%They are the first numerical experiments to show that $v_\mathrm{esc}$ determines the orbital architecture of planetary systems.

%% file: summary.tex
\section{Summary and Discussion}
\label{sec:summary}

% Summary

This study aims to obtain the fundamental scaling laws for the orbital architecture of planetary systems self-organized by gravitational scattering and collision among protoplanets.
We have investigated the orbital architecture of planetary systems formed from protoplanet systems by giant impacts using {\em N}-body simulations. 
We systematically changed the mass and location of the initial protoplanet systems and investigated their effects on the final planetary systems.  
%The basic orbital evolution of protoplanets during the giant impact phase is (1) an increase in orbital eccentricity and inclination due to gravitational scattering and (2) an increase in orbital spacing and decay in orbital eccentricity and inclination due to collisions. 
We found that the mean orbital separation and eccentricity of planetary systems normalized by the Hill radius are nearly independent of the total mass of the initial protoplanet systems, which means that the Hill radius scales the orbital architecture.
On the other hand, they show a positive dependence on the system semimajor axis.
This dependence can be scaled by the eccentric distance for the eccentricity given by the ratio between the two-body surface escape velocity and the Keplerian circular velocity $r_\mathrm{K}$.
We showed that the orbital architecture of planetary systems formed by giant impacts is scaled by $r_\mathrm{K}$, which consists of the Hill radius and the Hill-to-physical radius ratio.
In other words, $r_\mathrm{K}$ is an appropriate gravitational and collisional dynamics metric.
In units of $r_\mathrm{K}$, the mean orbital separation and the epicycle amplitude of a planetary system are $2.2\pm0.3$ and $0.3\pm0.1$, respectively.
%Idealization provides a context for further numerical study of these processes (Scheeres).
%This analysis with idealized models provides a context for further studies with more realistic models.

% Discussion

% planet mass
%Given the orbital separation, the planet mass is estimated by the isolation mass in the same way as the oligarchic growth \citep{1987Icar...69..249L,1998Icar..131..171K,2014ApJ...795L..15S,2023ASPC..534..863W} as
%\begin{equation}
%M \simeq 2\pi a b \Sigma = 0.36\scale{\Sigma}{1000\, \mathrm{gcm}^{-2}}{3/2} \scale{\tilde{b}_\mathrm{K}}{2}{3/2} \scale{\rho}{3\, \mathrm{gcm}^{-3}}{1/4} \scale{a}{0.1\, \au}{15/4} \scale{M_*}{M_\odot}{3/4} M_\oplus.
%\end{equation}
%This mass is generally consistent with the results of {\em N}-body simulations, though they have dispersion.
%Note that the growth mode of planets here is not oligarchic but rather cannibalism among protoplanets.
%The next paper will discuss the final mass of planets in more detail.
%E: M_tot、 a_Mの依存性について述べる?

% Comparision with observation
%E: 太陽系とCISEと整合的
%E: さやえんどうと整合的
%E: さやえんどうはaが大きいと弱まる
This scaling generally agrees with the observations.
In this model, the separation, eccentricity, and inclination of planetary orbits increase with their semimajor axis.
The orbital separations at $\simeq 0.1$ au and 1 au are consistent with those of the close-in super-Earths and the solar system terrestrial planets, respectively.
Furthermore, this model also generally agrees with the peas-in-a-pod pattern.
The orbital separation and mass deviations increase with the semimajor axis.
In other words, the system uniformity diminishes with increasing the semimajor axis.
These results mean that the peas-in-a-pod pattern diminishes with increasing the semimajor axis.

%E: resonant system への応用
This scaling could be applied to systems of protoplanets initially in resonant chains. 
This is because the initial resonant configuration breaks down when the giant impact phase begins.
%E: k = 1 になる限界について
Note that this model is only valid for $k \lesssim 1$.
For $k \gtrsim 1$, planet ejection may occur \citep[e.g.,][]{2020A&A...642A..23M}, which can modify the scaling.
%However, even in this case, the architecture of the stabilized planetary system after ejection is expected to follow this scaling, which is supported by the reinspection of the results in \citep[e.g.,][]{2020A&A...642A..23M}.
%However, even in this case, the structure of the stabilized planetary system after ejection is expected to follow this scaling, which is supported by a reanalysis of the results in \citep[e.g.,][]{2020A&A...642A..23M}.
%This may modify the scaling, which will be the subject of another paper.
%E: 最小のbとeを決める?
%E: (永年)長期進化について?

% Future plan
The present analysis with idealized models provides a context for further studies with more realistic models.
In a subsequent paper, we will further explore the parameter dependence of the orbital architecture to confirm the universality of the scaling.
We also plan to investigate the effects of initial resonant configurations and planet ejection on the orbital architecture.
%We also plan to compare our model with the observations of close-in super-Earth systems.

%% file: acknowledgement.tex
E.K.\ is supported by JSPS KAKENHI Grants No.\ 18H05438, 24K00698 and 24H00017.
R.S.\ is partially supported by ISF, MOS, and NSF/BSF, and GIF grants.
Numerical computations were partially conducted on the PC cluster at the Center for Computational Astrophysics, National Astronomical Observatory of Japan.
This research was supported in part by the Munich Institute for Astro-, Particle and BioPhysics (MIAPbP) funded by the Deutsche Forschungsgemeinschaft (DFG, German Research Foundation) under Germany's Excellence Strategy – EXC-2094 – 390783311 and by grant NSF PHY-2309135 to the Kavli Institute for Theoretical Physics (KITP).